\documentclass{appolb}
\usepackage{epsfig}

\begin{document}

\title{Exact Results for Spectra of Overdamped Brownian Motion in Fixed
and Randomly Switching Potentials
\thanks{Presented at the XVI Marian Smoluchowski Symposium on Statistical Physics, \ \\
Zakopane, Poland, September 6-11, 2003.}}
\author{A.A. Dubkov, V.N. Ganin
\address{Radiophysics Department, Nizhni Novgorod State
University, 23 Gagarin Ave., 603950 Nizhni Novgorod, Russia}\and
B. Spagnolo
\address{INFM-Group of Interdisciplinary Physics and  Dipartimento di
Fisica e Tecnologie Relative, Universit\`a di Palermo, Viale delle
Scienze - 90128 Palermo, Italy} } \maketitle

\begin{abstract}
The exact formulae for spectra of equilibrium diffusion in a fixed
bistable piecewise linear potential and in a randomly flipping
monostable potential are derived. Our results are valid for
arbitrary intensity of driving white Gaussian noise and arbitrary
parameters of potential profiles. We find: (i) an exponentially
rapid narrowing of the spectrum with increasing height of the
potential barrier, for fixed bistable potential; (ii) a nonlinear
phenomenon, which manifests in the narrowing of the spectrum with
increasing mean rate of flippings, and (iii) a nonmonotonic
behaviour of the spectrum at zero frequency, as a function of the
mean rate of switchings, for randomly switching potential. The
last feature is a new characterization of resonant activation
phenomenon.
\end{abstract}

\section{Introduction}

Spectral densities of fluctuations provide an important tool to
characterize physical systems, because they can be measured
directly in experiments. The investigations of spectra are useful
to observe and analyze the interplay between fluctuations,
relaxation and nonlinearity which are inherent to real physical
systems. This interplay ranks among the most challenging problems
of modern nonlinear physics and forms the basis of well-known
nonlinear phenomena like stochastic resonance \cite{Gam}, resonant
activation \cite{Doe}, noise-enhanced stability
\cite{Man,Agu}, ratchet-effect \cite{Mag,Rei}, etc. \\
\indent The exact formulae for spectra of fluctuations in
nonlinear dynamical systems were first derived for thermal
diffusion in fixed potentials. Caughey and Dienes \cite{Cau}
pioneered in applying analytical method based on Laplace transform
of conditional probability density to the first-order system with
$V$-shaped potential. Another approach has its origins in the
expansion of probability density of transitions in terms of
Fokker-Planck kinetic operator eigenfunctions. This method was
applied in \cite{Mor} for obtaining correlation function of a
bistable system with rectangular potential profile. We would also
mention theoretical and numerical calculations reported in
refs.~\cite{Dyk}, concerning the spectra of underdamped
double-well system driven by white Gaussian noise. In these papers
the spectral peaks corresponding to standard resonance and
transitions between steady states have been revealed. Stationary
spectra of fluctuations for monostable and bistable potential
profiles, by analog simulations of underdamped stochastic system
driven by colored noise, have been experimentally obtained in
ref.~\cite{Mar}. The model of one-dimensional Brownian motion in
singular potential like the potential of hydrogen atom was
investigated in \cite{Ouy}. Authors detected some region of power
spectrum with $1/f$ frequency dependence.\\
\indent Despite a lot of work has been done to analyze spectra of
fluctuations in the presence of one noise source, there is however
lack of investigation on the so-called two-noise system spectra of
fluctuations. A paradigmatic model is the overdamped Brownian
motion in a randomly fluctuating potential. This model is being
studied intensively in view of wide application in physics,
chemistry and biology. However, an exact analytical results have
been obtained only for escape rates, as mean first-passage times
and lifetimes \cite{Agu,Agu2} and stationary probability
distributions of Brownian motion \cite{Dub-2}. In this paper we
report the exact calculations of diffusion spectrum for Brownian
particle moving in fixed double-well potential and dichotomously
switching linear potential. Our theoretical results, based on
Markovian theory and on Laplace transform of conditional
probability density, are valid for arbitrary intensity of driving
white Gaussian noise and arbitrary parameters of potential
profiles. We find: (i) a narrowing of the spectrum with increasing
height of the potential barrier for fixed potential; (ii) a
narrowing of the spectrum with increasing mean rate of flippings,
and (iii) a nonmonotonic behaviour of the spectrum at zero
frequency, as a function of the mean rate of switchings, for
randomly switching potential. This last behaviour is a new
characterization of resonant activation phenomenon \cite{Doe}.

\section{Basic equations}

Let us consider an overdamped Brownian motion in a fixed potential
$U(x)$ described by Langevin equation
\begin{equation}
\frac{dx}{dt}=-\frac{dU\left(  x\right)  }{dx}+\xi\left(  t\right)
,\label{Lang}%
\end{equation}
where $x(t)$ is the position of Brownian particle, $\xi(t)$ is a
$\delta$-correlated Gaussian noise with zero mean and intensity
$2D$. The Fokker-Planck equation, or Smoluchowski equation
\cite{Ris}, for the conditional probability density
$W(x,t\left\vert x_{0},0\right.)$ of Markovian random process
$x(t)$, corresponding to (\ref{Lang}), is
\begin{equation}
\frac{\partial W}{\partial t}=\frac{\partial}{\partial x}\left[
\frac{dU\left(  x\right)  }{dx}W\right]
+D\frac{\partial^{2}W}{\partial x^{2}}
\label{F-P}
\end{equation}
with initial condition
\begin{equation}
W(x,0\left\vert x_{0},0\right.)=\delta\left( x-x_{0}\right).
\label{Ini}
\end{equation}
Let us assume that a stationary regime exists, then the
probabilistic flow equals zero at $x\rightarrow
\pm\infty$
\begin{equation}
\left[  D\frac{\partial W}{\partial x}+U^{\prime }\left(  x\right)
W\right] _{x=\pm\infty}=0.\label{Krai}%
\end{equation}
The correlation function of Brownian particle displacement $x(t)$
in a stationary state can be calculated as \cite{Dub}
\begin{equation}
K\left[  \tau\right]
=\int\limits_{-\infty}^{+\infty}x_{0}W_{\infty}\left( x_{0}\right)
dx_{0}\int\limits_{-\infty}^{+\infty}xW(x,\tau\left\vert
x_{0},0\right.  )dx,
\label{CF}
\end{equation}
where $W_{\infty}(x)$ is the stationary probability density (SPD)
\cite{Ris,Stra}
\begin{equation}
W_{\infty}\left(x\right)= C\cdot \e ^{-U\left(  x\right)
/D},\qquad C= \left[ \int\limits_{-\infty}^{+\infty}\e ^{-U\left(
x\right) /D}dx\right] ^{-1}. \label{SPD}
\end{equation}
To obtain the correlation function $K\left[\tau\right]$ we need to
solve the second-order partial differential equation (\ref{F-P})
using eigenfunction expansion \cite{Mor,Ris}. However, as shown in
\cite{Cau,Dub}, the determination of SPD (\ref{SPD}) together with
the Laplace transform method are sufficient for calculating the
spectral density. In fact from Wiener-Khinchin theorem we have
\begin{equation}
S\left(  \omega\right)  =\frac{1}{2\pi}\int\limits_{-\infty}^{+\infty}K\left[
\tau\right]  \cos\left(  \omega\tau\right)  d\tau=\frac{1}{\pi}%
\mathrm{Re}\left\{  \tilde{K}\left[ i\omega\right] \right\}
,\label{Spec}%
\end{equation}
where $\tilde{K}\left[  p\right]$ is Laplace transform of $K\left[
\tau\right]$. By Laplace transforming (\ref{F-P}), with initial
condition (\ref{Ini}), we obtain
\begin{equation}
D\frac{d^{2}Y}{dx^{2}}+\frac{d}{dx}\left[  U^{\prime }\left(
x\right) Y\right]  -pY=-\delta\left(  x-x_{0}\right)  ,\label{Lap-w}%
\end{equation}
\ie a second-order ordinary differential equation, where $Y\left(
x,x_{0},p\right)$ is the Laplace transform of conditional
probability density
\begin{equation}
Y\left(  x,x_{0},p\right) =\int\limits_{0}^{+\infty}\e
^{-pt}W(x,t\left\vert x_{0},0\right.  )
dt.\label{Lapl}%
\end{equation}
According to Eqs. (\ref{Krai}) and (\ref{Lapl}) we solve
(\ref{Lap-w}) with boundary conditions
\begin{equation}
\left[  D\frac{dY}{dx}+U^{\prime}\left(  x\right)  Y\right]
_{x=\pm\infty }=0.
\label{Boun}
\end{equation}
By using Eqs. (\ref{CF}) and (\ref{Lapl}) the Laplace transform
$\tilde{K}\left[  p\right]  $ of the correlation function is
\begin{equation}
\tilde{K}\left[  p\right]
=\int\limits_{-\infty}^{+\infty}x_{0}W_{\infty }\left(
x_{0}\right)  dx_{0}\int\limits_{-\infty}^{+\infty}xY\left(
x,x_{0},p\right)  dx.
\label{Lap-cor}
\end{equation}
Then, after substitution of $p=i\omega$ in (\ref{Lap-cor}) we can
find the spectral density $S(\omega)$ from (\ref{Spec}). Thus for
calculating spectrum it will suffice to solve ordinary
differential equation (\ref{Lap-w}) and make double integration,
instead of solving partial differential equation (\ref{F-P}). To
end we need the explicit expression of the internal integral in
(\ref{Lap-cor}). By multiplying both parts of (\ref{Lap-w}) on $x$
and integrating it over the total area, using boundary conditions
(\ref{Boun}), we obtain
\begin{eqnarray}
&& G\left(  x_{0},p\right)
\equiv\int\limits_{-\infty}^{+\infty}xY\left( x,x_{0},p\right)
dx=\frac{x_{0}}{p}-\frac{D}{p}\left[  Y\left(  \infty
,x_{0},p\right)  -Y\left(  -\infty,x_{0},p\right)  \right] \nonumber\\
&& -\frac{1}{p}\int\limits_{-\infty}^{+\infty}U^{\prime}\left(
x\right)
Y\left(  x,x_{0},p\right)  dx.\label{Prel}%
\end{eqnarray}

\section{Fixed bistable potential}

Let us calculate the spectral density for symmetric double-well
piecewise linear potential (see Fig.~1)
\begin{equation}
U(x)=\left\{
\begin{array}
[c]{cc}%
E\left(  1-\left\vert x\right\vert \left/  L\right.  \right)  , &
\left\vert
x\right\vert \le L,\\
+\infty,\quad & \left\vert x\right\vert >L.
\end{array}
\right. \label{Pot}%
\end{equation}
\begin{figure}[htbp]
\begin{center}
\includegraphics[width=5cm]{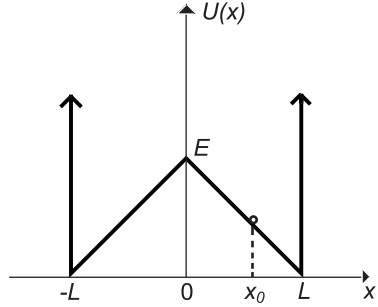}
\end{center}
\caption{\small \emph{Double-well piecewise linear
potential.}\bigskip} \label{beta_1}
\end{figure}
Substituting (\ref{Pot}) in (\ref{Lap-w}),(\ref{Boun}) we obtain
the following equation for the Laplace transform $Y(x,x_{0},p)$ of
conditional probability density
\begin{equation}
DY^{\prime\prime}-\frac{E}{L}\left[  \mathrm{sgn}\left( x\right)
Y\right]  ^{\prime
}-pY=-\delta\left(  x-x_{0}\right) \label{Lap-bi}%
\end{equation}
with the conditions at reflecting boundaries $x=\pm L$
\begin{equation}
\left[  DY^{\prime}-\frac{E}{L}\mathrm{sgn}\left(  x\right)
Y\right]  _{x=\pm
L}=0,\label{Boun-bi}%
\end{equation}
where $\mathrm{sgn}(x)$ is the sign function. Because of
normalization condition for $Y(x,x_{0},p)$
\[
\int\limits_{-L}^{L}Y\left(  x,x_{0},p\right) dx=\frac{1}{p},
\]
equation (\ref{Prel}) gives
\begin{eqnarray}
&& G\left(  x_{0},p\right)  =\frac{x_{0}}{p}+\frac{E}{p^{2}L}-\frac{D}%
{p}\left[  Y\left(  L,x_{0},p\right)  -Y\left(  -L,x_{0},p\right)  \right]
\nonumber\\
&& -\frac{2E}{pL}\int\limits_{-L}^{0}Y\left(  x,x_{0},p\right)
dx.\label{F-int}%
\end{eqnarray}

To derive the function $Y(x,x_{0},p)$ we consider first $x_{0}>0$
and solve homogeneous equation (\ref{Lap-bi}) in regions $-L\le x
\le 0$, $0 \le x \le x_{0}$, $x_{0} \le x \le L$ separately. Then
we apply the continuity conditions at the points $x=0$ and
$x=x_{0}$
\begin{eqnarray}
&& D\left[  Y^{\prime}\left(  +0,x_{0},p\right)  -Y^{\prime}\left(
-0,x_{0},p\right)  \right] -\frac{E}{L}\left[ Y\left(
+0,x_{0},p\right)  +Y\left(  -0,x_{0},p\right) \right]
=0,\nonumber\\
&& D\left[  Y^{\prime}\left(  x_{0}+0,x_{0},p\right)
-Y^{\prime}\left(
x_{0}-0,x_{0},p\right)  \right]  =-1,\nonumber\\
&& Y\left(  +0,x_{0},p\right)  =Y\left(  -0,x_{0},p\right)  ,\nonumber\\
&& Y\left(  x_{0}+0,x_{0},p\right)  =Y\left(
x_{0}-0,x_{0},p\right)
.\label{Conti}%
\end{eqnarray}
Solving (\ref{Lap-bi}) in above-mentioned regions and taking into
account the boundary conditions (\ref{Boun-bi}) we arrive at
\begin{equation}
Y\left(  x,x_{0},p\right)  =\left\{
\begin{array}
[c]{ll} c_{1}\left[ \e^{-\lambda_{1}\left(  x+L\right)  }-\left(
\lambda_{2}\left/ \lambda_{1}\right.  \right)
\e^{-\lambda_{2}\left(x+L\right)}\right] ,
-L\le x \le 0,\\
c_{2}\cdot \e^{\lambda_{1}x}+c_{3}\cdot \e^{\lambda_{2}x},
\qquad \qquad \qquad \qquad 0 \le x \le x_{0},\\
c_{4}\left[ \e^{\lambda_{1}\left(  x-L\right)  }-\left(
\lambda_{2}\left/ \lambda_{1}\right.  \right)
\e^{\lambda_{2}\left( x-L\right)  }\right]  , \qquad x_{0} \le x
\le L,
\end{array}
\right. \label{Solu}
\end{equation}
where $\lambda_{1,2}=\left(E\pm\sqrt{E^{2}+4pDL^{2}}\right) \left/
\left( 2DL\right)\right.$. Substitution of (\ref{Solu}) in
(\ref{F-int}) gives
\begin{eqnarray}
&& G\left(  x_{0},p\right)  =\frac{x_{0}}{p}+\frac{E}{p^{2}L}+\frac{D}%
{p}\left(  1-\frac{\lambda_{2}}{\lambda_{1}}\right)  \left(  c_{1}%
-c_{4}\right) \nonumber\\
&& +\frac{2E}{pL\lambda_{1}}c_{1}\left( \e^{-\lambda_{1}L}-\e^{-\lambda_{2}%
L}\right)  .\label{Fun-G}%
\end{eqnarray}
Calculating unknown constants $c_{1}$ and $c_{4}$ from the continuity
conditions (\ref{Conti}) and substituting theirs in (\ref{Fun-G}) we have
\begin{eqnarray}
&& G\left(  x_{0},p\right) =\frac{x_{0}}{p}+\frac{E}{p^{2}L}+\frac
{\e^{-\lambda_{1}x_{0}}}{p^{2}L}\cdot\frac{pL+E\lambda_{2}\e^{-\lambda_{2}L}
}{\lambda_{1}\e^{-\lambda_{1}L}-\lambda_{2}\e^{-\lambda_{2}L}}\nonumber\\
&& +\frac{\e^{-\lambda_{2}x_{0}}}{p^{2}L}\cdot\frac{pL+E\lambda_{1}%
\e^{-\lambda_{1}L} }{\lambda_{2}\e^{-\lambda_{2}L}-\lambda_{1}\e^{-\lambda_{1}L}%
}.\label{Fin-G}%
\end{eqnarray}

To obtain the function $G\left(  x_{0},p\right)  $ in the region
$x_{0}<0$ we use symmetry considerations. Because of the symmetry
of the potential $U(x)$, the SPD (\ref{SPD}) is an even function
of $x$, $W(x,t\left\vert x_{0},0\right.) = W(-x,t\left\vert
-x_{0},0\right.)$ and $Y\left( x,x_{0},p\right) =Y\left(
-x,-x_{0},p\right)$. So $G\left( x_{0},p\right)$ is an odd
function of variable $x_0$: $G\left( x_{0},p\right) =-G\left(
-x_{0},p\right)$, and from (\ref{SPD}),(\ref{Lap-cor}) we obtain
\begin{equation}
\tilde{K}\left[  p\right]  =\frac{\beta}{\left( \e^{\beta}-1\right)  L}%
\int\limits_{0}^{L}x_{0}G\left(  x_{0},p\right) \e^{\beta x_{0}/L}%
dx_{0}, \label{Int-prel}
\end{equation}
where $\beta=E\left/ D\right.$ is the dimensionless height of
potential barrier. Substitution of (\ref{Fin-G}) in
(\ref{Int-prel}) and subsequent integration gives the following
result for Laplace transform of correlation function in stationary
state
\begin{eqnarray}
&& \tilde{K}\left[  p\right]  =\frac{\left\langle x^{2}\right\rangle }{p}%
-\frac{D}{p^{2}}+\frac{\beta D}{p^{2}\left(  1-\e^{-\beta}\right)
\left( \alpha_{1}\e^{\alpha_{2}}-\alpha_{2}\e^{\alpha_{1}}\right)  }\nonumber\\
&& \times\left\{ \e^{\alpha_{1}}-\e^{\alpha_{2}}+4\beta\left[
\frac{\sinh ^{2}\left(  \alpha_{2}\left/  2\right.  \right)
}{\alpha_{2}}-\frac{\sinh ^{2}\left(  \alpha_{1}\left/  2\right.
\right)  }{\alpha_{1}}\right] \right\}  , \label{Cor-fun}
\end{eqnarray}
where $\alpha_{1,2}=\left( \beta\pm\sqrt{\beta^{2}+4pL^{2}\left/
D\right. }\right) /2$. To obtain the spectral density of
coordinate fluctuations of Brownian particle moving in a
double-well potential (\ref{Pot}) it remains to put in
(\ref{Cor-fun}) $p=i\omega$ and find its real part. However, we
will not report here the exact formula for the spectrum because of
its complicated expression. We give here the spectrum for
particular case of a rectangular potential well, \ie in the
absence of a barrier $\left( \beta=0\right)$. We have
$\left\langle x^{2}\right\rangle =L^{2}\left/ 3\right.$ and from
(\ref{Cor-fun}) we get
\begin{equation}
\tilde{K}\left[  p\right]  =\frac{L^{2}}{3p}+\frac{D}{p^{2}}\left(
\frac{\tanh L\sqrt{p\left/  D\right.  }}{L\sqrt{p\left/  D\right.  }%
}-1\right),
\label{Simp}
\end{equation}
so after substitution of $p=i\omega$ in (\ref{Simp}) we arrive
finally at $\left(  \omega>0\right)$
\begin{equation}
S\left(  \omega\right)  =\frac{D}{\pi\omega^{2}}\left(  1-\frac{1}{L}%
\sqrt{\frac{D}{2\omega}}\cdot\frac{\sinh L\sqrt{2\omega\left/
D\right. }+\sin L\sqrt{2\omega\left/  D\right.  }}{\cosh
L\sqrt{2\omega\left/ D\right.  }+\cos L\sqrt{2\omega\left/
D\right.  }}\right).
\label{Rec-well}
\end{equation}

\section{Discussions}

Spectral densities of Brownian diffusion, obtained from
(\ref{Cor-fun}), for different values of the potential barrier
height are plotted in Fig.~2.
\begin{figure}[htbp]
\begin{center}
\includegraphics[width=6cm]{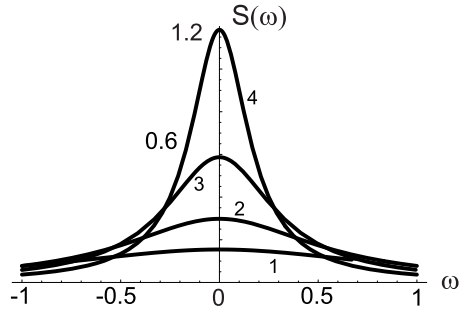}
\end{center}
\caption{\small \emph{Spectral density of Brownian particle
displacement for different values of dimensionless height ($\beta
= E/D$) of potential barrier: curve $1$ - $\beta$ $=2$, curve $2$
- $\beta=3$, curve $3$ - $\beta$ $=4$, curve $4$ - $\beta$ $=5$.
Parameters are $L=1$, $D=1$.}} \label{beta_2}
\end{figure}
As shown in Fig.~2, the spectral density has a maximum at zero
frequency, which is a general property of Markovian random
processes. We see also that the spectrum $S(\omega)$ narrows very
rapidly with increasing height of potential barrier, and its value
at zero frequency increases fast. This nonlinear phenomenon is due
to very rare transitions between steady states when the barrier is
high with respect to the noise intensity \cite{Kra}. Brownian
particles therefore move within a potential well for most of the
time, and their displacements vary very slowly. As a result, the
width of spectral density decreases.

To verify this hypothesis we compare the behaviours of the
spectral width and the mean rate of transitions between steady
states as a function of potential barrier height. First we find
the value of spectral density at zero frequency $S(0)=$
$\tilde{K}\left[ 0\right] \left/ \pi\right.$. Let us expand the
function (\ref{Cor-fun}) in power series on small parameter
$\alpha_{2}$. Then we express this parameter in terms of small
parameter $p$
\[
\alpha_{2}\simeq-\frac{pL^{2}}{\beta D}+\frac{p^{2}L^{4}}{\beta^{3}D^{2}},
\]
and after calculation of the limit $p\rightarrow0$, we get
\begin{equation}
S(0)=\frac{L^{4}}{\pi D}\cdot\frac{\left(  \beta-1\right)
^{2}\e^{2\beta }-\left(  \beta^{3}-3\beta^{2}+4\beta-4\right)
\cdot \e^{\beta}-5}{\beta ^{4}\left( \e^{\beta}-1\right)}.
\label{Zero}
\end{equation}
The value $S(0)$ increases therefore as an exponential law
$S(0)\sim \e^{\beta}\left/ \beta^{2} \right. $, with increasing
height of potential barrier $\beta$ and takes the finite value
$2L^{4}\left/ \left( 15\pi D\right)  \right.$ for $\beta=0$. This
value corresponds to a diffusion in rectangular potential well
(see (\ref{Rec-well})). The width of the spectral density with a
maximum at zero frequency can be defined as \cite{Stra}
\begin{equation}
\Pi=\int\limits_{0}^{+\infty}S(\omega)d\omega\left/  S(0)\right.
=\frac{\left\langle x^{2}\right\rangle }{2S(0)}. \label{Def}
\end{equation}
The variance of Brownian particle position in stationary state
from Eqs. (\ref{SPD}) and (\ref{Pot}) is
\begin{equation}
\left\langle x^{2}\right\rangle =\frac{L^{2}\left[  \left(  \beta^{2}%
-2\beta+2\right)  \cdot \e^{\beta}-2\right]  }{\beta^{2}\left(
\e^{\beta }-1\right)  } \label{Var}
\end{equation}
and increases monotonically from the value $L^{2}\left/ 3\right.
$, which takes for $\beta\rightarrow0$, to the value $L^{2}$,
which takes for $\beta\rightarrow\infty$, due to the finite area
of diffusion. Substituting Eqs. (\ref{Zero}) and (\ref{Var}) in
Eq. (\ref{Def}) we obtain
\begin{equation}
\Pi=\frac{\pi D}{2L^{2}}\cdot\frac{\beta^{2}\left[  \left(  \beta^{2}%
-2\beta+2\right)  \cdot \e^{\beta}-2\right]  }{\left(
\beta-1\right) ^{2}\e^{2\beta}-\left(
\beta^{3}-3\beta^{2}+4\beta-4\right)  \cdot \e^{\beta }-5}.
\label{Width}
\end{equation}
By introducing correlation time similar to (\ref{Def})
\[
\tau_{c}=\int\limits_{0}^{+\infty}K\left[ \tau \right] d\tau
\left/ K\left[ 0\right] \right.
\]
we find from Eqs. (\ref{Spec}) and (\ref{Def})
\[
\tau_{c}=\frac{\pi S\left( 0\right) }{\left\langle
x^{2}\right\rangle }=\frac{\pi }{2\Pi },
\]
and equation (\ref{Width}) gives the exact correlation time for
bistable potential, recently obtained in \cite{Dub}. The spectral
width decreases monotonically with increasing height of potential
barrier from the value $5\pi D\left/  \left( 4L^{2}\right) \right.
$, taken for $\beta\rightarrow0$, to zero, taken for
$\beta\rightarrow\infty$.

The mean rate of transitions, from one stable state to the other,
can be determined through the mean first passage time (MFPT) to
reach the top of barrier ($x=0$) from the bottom of well ($x=L$),
by solving the following differential equation \cite{Ris,Stra}
($x>0$)
\[
D\tau^{\prime\prime}\left(  x\right)  +\frac{E}{L}\tau^{\prime}\left(
x\right)  =-1
\]
with boundary conditions: $\tau^{\prime}\left(L\right) =0$,
$\tau\left(0\right)=0$. After simple calculations we get for
$\tau(L)$
\[
\tau\left(L\right)
=\frac{L^{2}}{D}\cdot\frac{\e^{\beta}-1-\beta}{\beta^{2}},
\]
and for mean rate of transitions between two steady states
\begin{equation}
\Omega =\frac{D}{L^2}\cdot\frac{\beta^2}{\e^{\beta}-1-\beta}.
\label{Trans}
\end{equation}
\begin{figure}[htbp]
\begin{center}
\includegraphics[width=5cm]{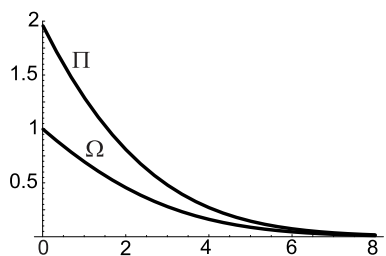}
\end{center}
\caption{\small \emph{The spectral width and the mean rate of
transitions between two steady states vs the dimensionless height
of potential barrier. Parameters are $L=1$,
$D=0.5$.}}\label{beta_3}
\end{figure}

In Fig.~3 we report the behaviours of $\Pi$ and $\Omega$ as a
functions of dimensionless height of potential barrier $\beta$.
The curves expressed by Eqs. (\ref{Width}) and (\ref{Trans})
practically coincide at large values of $\beta$. Thus, the mean
rate of transitions is approximately the spectral width of
Brownian particle coordinate fluctuations in a stationary state.

\section{Randomly switching monostable potential}

Let us consider now two-noise nonlinear system, namely,
one-dimensional overdamped Brownian motion in a fluctuating
potential described by the following Langevin equation
\begin{equation}
\frac{dx}{dt}=-\frac{\partial\Phi\left(  x,t\right)  }{\partial
x}+\xi\left( t\right) ,
\label{New-Lang}
\end{equation}
where $\xi(t)$ is white Gaussian noise with zero mean and
intensity $2D$, $\Phi(x,t)=U(x)+a\eta(t)x$,  $U(x)$ is the same
potential (\ref{Pot}) but without barrier $(E=0)$, and $\eta(t)$
is Markovian dichotomous noise switching with mean rate $\nu$
between the values $\pm1$. In other words, we analyze Brownian
diffusion in monostable potential with two randomly switching
stable states near reflecting boundaries at $x=\pm L$ (see
Fig.~4).
\begin{figure}[htbp]
\begin{center}
\includegraphics[height=4cm,width=4.5cm]{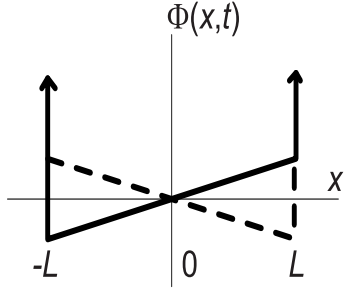}
\end{center}
\caption{\small \emph{Randomly switching monostable
potential.}\bigskip} \label{beta_4}
\end{figure}

Let us rewrite for our case the closed set of differential
equations for probability density $W(x,t)$, recently obtained in
\cite{Dub-2}, in the diffusion interval $\left( -L,L\right)$
\begin{eqnarray}
&& \frac{\partial W}{\partial t}=a\frac{\partial Q}{\partial
x}+D\frac
{\partial^{2}W}{\partial x^{2}},\nonumber\\
&& \frac{\partial Q}{\partial t}=-2\nu Q+a\frac{\partial W}{\partial x}%
+D\frac{\partial^{2}Q}{\partial x^{2}}
\label{Two-equa}
\end{eqnarray}
with the following conditions at reflecting boundaries
\begin{equation}
\left( DW^{\prime}+aQ\right) _{x=\pm L}=0,\qquad\left( DQ^{\prime
}+aW\right) _{x=\pm L}=0, \label{Boun-WQ}
\end{equation}
where $Q(x,t)=\left\langle \eta(t)\delta(x-x(t))\right\rangle =W\left(
x,t\right) \left\langle \eta(t)\left\vert x\left(  t\right)
=x\right. \right\rangle$ is an auxiliary function \cite{Dub-2}. We use the same method as
for fixed potential.

According to Eqs. (\ref{Two-equa}) and (\ref{Boun-WQ}) and initial
conditions for the functions $W(x,t)$, $Q(x,t)$:
$W(x,0)=\delta(x-x_{0})$, $Q(x,0)=0$, we solve the following
system of differential equations in the interval $(-L,L)$
\begin{eqnarray}
&& DY^{\prime\prime}+aZ^{\prime}-pY=-\delta(x-x_{0}),\nonumber\\
&& DZ^{\prime\prime}+aY^{\prime}-\left(  p+2\nu\right)Z=0
\label{Y-Z}
\end{eqnarray}
with boundary conditions
\begin{equation}
\left( DY^{\prime}+aZ\right) _{x=\pm L}=0,\qquad\left( DZ^{\prime
}+aY\right) _{x=\pm L}=0. \label{New-boun}
\end{equation}
Here $Y(x,x_{0},p)$ and $Z(x,x_{0},p)$ are the Laplace transforms
of conditional probability density and of auxiliary function
respectively. By putting $E=0$ in (\ref{F-int}) we get
\begin{equation}
G\left(  x_{0},p\right)  =\frac{x_{0}}{p}-\frac{D}{p}\left[
Y\left( L,x_{0},p\right)  -Y\left(  -L,x_{0},p\right)  \right].
\label{G-x}
\end{equation}
Now we solve the homogeneous set of linear differential equations
(\ref{Y-Z}) in two regions: $-L\le x\le x_0$ and $x_0\le x\le L$.
Then we find eight unknown constants from the boundary conditions
(\ref{New-boun}) and continuity conditions at the point $x=x_{0}$
\begin{eqnarray}
&& Y\left\vert _{x=x_{0}-0}\right.  =Y\left\vert
_{x=x_{0}+0}\right.  ,\qquad Y^{\prime}\left\vert
_{x=x_{0}-0}\right. =Y^{\prime}\left\vert _{x=x_{0}+0}\right.
+1\left/
D\right.  ,\nonumber\\
&& Z\left\vert _{x=x_{0}-0}\right.  =Z\left\vert
_{x=x_{0}+0}\right.  ,\qquad
Z^{\prime}\left\vert _{x=x_{0}-0}\right.  =Z^{\prime}\left\vert _{x=x_{0}+0}\right.  .\nonumber%
\end{eqnarray}
After some algebra we obtain from (\ref{G-x})
\begin{eqnarray}
&& G\left(  x_{0},p\right)  =\frac{x_{0}}{p}\label{Full-G}\\
&& -\frac{\left(  D\rho_{1}^{2}-p\right)  \sinh\rho_{1}L\sinh\rho_{2}%
x_{0}-\left(  D\rho_{2}^{2}-p\right)  \sinh\rho_{2}L\sinh\rho_{1}x_{0}%
}{p\left[  \rho_{2}\left(  D\rho_{1}^{2}-p\right)  \sinh\rho_{1}L\cosh\rho
_{2}L-\rho_{1}\left(  D\rho_{2}^{2}-p\right)  \sinh\rho_{2}L\cosh\rho
_{1}L\right]  },\nonumber
\end{eqnarray}
where
\begin{equation}
\rho_{1,2}=\sqrt{\frac{\gamma^{2}}{2}+\frac{p}{D}\pm\sqrt{\frac{\gamma^{4}}%
{4}+\frac{pa^{2}}{D^{3}}}},\qquad\gamma=\sqrt{\frac{a^{2}}{D^{2}}+\frac{2\nu
}{D}}.\label{Ro}%
\end{equation}

To find the Laplace transform (\ref{Lap-cor}) of correlation
function in stationary regime we use the expression of SPD for our
system, derived in ref.~\cite{Dub-2},
\begin{equation}
W_{\infty}\left(  x\right)
=\frac{1}{2L}\cdot\frac{1+\mu\cosh\gamma x\left/ \cosh\gamma
L\right.  }{1+\mu\tanh\gamma L\left/  \left(  \gamma L\right)
\right.  },
\label{New-SPD}
\end{equation}
where $\mu=a^{2}\left/  (2\nu D)\right.$. After substitution of
Eqs. (\ref{Full-G}) and (\ref{New-SPD}) into Eq. (\ref{Lap-cor})
and integration we get
\begin{eqnarray}
&& \tilde{K}\left[  p\right]  =\frac{\left\langle x^{2}\right\rangle }{p}%
-\frac{1}{p\left[  1+\mu\tanh\gamma L\left/  \left(  \gamma L\right)  \right.
\right]  }\nonumber \\
&& \times\frac{\left(  D\rho_{1}^{2}-p\right)  R\left(
\rho_{2}\right) \tanh\rho_{1}L-\left(  D\rho_{2}^{2}-p\right)
R\left(  \rho_{1}\right)
\tanh\rho_{2}L}{\rho_{2}\left(  D\rho_{1}^{2}-p\right)  \tanh\rho_{1}%
L-\rho_{1}\left(  D\rho_{2}^{2}-p\right)
\tanh\rho_{2}L},\label{Final-K}
\end{eqnarray}
where
\begin{eqnarray}
&& R\left(  z\right)  =\frac{1}{z}\left(  1-\frac{\tanh
zL}{zL}\right) +\frac{\gamma\mu\tanh zL}{z^{2}-\gamma^{2}}\left[
\frac{2z}{\left(
z^{2}-\gamma^{2}\right)  L}-\tanh zL\right] \nonumber\\
&& +\frac{z\mu}{z^{2}-\gamma^{2}}\left(  1-\frac{z^{2}+\gamma^{2}}{z^{2}%
-\gamma^{2}}\cdot\frac{\tanh zL}{zL}\right)  .
\end{eqnarray}
To obtain the exact formula for the spectral density of Brownian
particle position it remains to put in equation (\ref{Final-K})
$p=i\omega$ and find the real part of expression. In the absence
of flippings ($a=0$, $\mu = 0$) we find from Eq. (\ref{Ro}):
$\rho_{1}=\sqrt{\gamma^{2}+p\left/  D\right.}$,
$\rho_{2}=\sqrt{p\left/ D\right.  }$ and obtain the result for
rectangular potential well of equation (\ref{Simp}).

\section{New characterization of resonant activation}

The evolution of spectrum shape with varying switchings mean rate
$\nu$ is shown in Fig.~5. The spectral density of Brownian
diffusion in this non-Markovian case has also a maximum at zero
frequency.
\begin{figure}[htbp]
\begin{center}
\includegraphics[width=6cm]{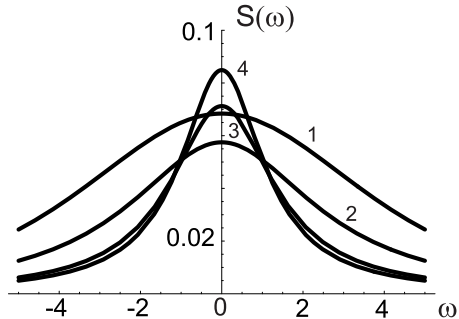}
\end{center}
\caption{\small \emph{Spectral density $S\left(\omega\right)$ for
different values of switchings mean rate $\nu$: curve $1$ -
$\nu=0.01$, curve $2$ - $\nu=3$, curve $3$ - $\nu=30$. The curve
$4$ corresponds to a free diffusion in rectangular potential well
$\left(  a=0\right)$. The parameter set is: $L=1$, $a=3$,
$D=0.5$.}} \label{beta_5}
\end{figure}
For very large values of $\nu$, the spectral density approximates
to the curve corresponding to a free diffusion in rectangular
potential well. The main feature of Fig.~5 is that the spectrum at
zero frequency $S\left(0\right)$ shows nonmonotonic behaviour with
increasing switchings mean rate $\nu$. Namely, $S\left(0\right)$
initially decreases, reaches a minimum and then increases reaching
asymptotically the value $2L^{4}\left/ \left(  15\pi D\right)
\right. $, obtained for rectangular potential well.

Let us find the analytical expression of $S\left(0\right)
=\tilde{K}\left[ 0\right]  \left/  \pi\right.$. Using the
approximate expressions for
parameters $\rho_{1}$, $\rho_{2}$ at small $p$ (see (\ref{Ro}))
\[
\rho_{1}\simeq\gamma+\frac{p}{2\gamma D}\left(  1+\frac{a^{2}}{\gamma^{2}%
D^{2}}\right)  ,\qquad\rho_{2}\simeq\frac{\sqrt{2\nu p}}{\gamma D}%
\]
and formula for the variance \cite{Dub-2}

\begin{equation}
\left\langle x^{2}\right\rangle =\frac{\gamma^{3}L^{3}+3\mu\left[
\left( 2+\gamma^{2}L^{2}\right) \tanh\gamma L-2\gamma L\right]
}{3\gamma ^{3}L\left[ 1+\mu\tanh\gamma L\left/  \left(  \gamma
L\right)  \right. \right]}
 \label{var2}
\end{equation}
we get from Eq. (\ref{Final-K})
\begin{eqnarray}
&& S\left(  0\right)  =\frac{1}{60\pi\gamma^{6}D^{3}\left[
1+\mu\tanh\gamma L\left/ \left( \gamma L\right) \right.
\right]}\{ 16\nu D\gamma^{4}L^{4} \nonumber \\
&& +5a^{2}[ 60+27\mu+4\gamma^{2}L^{2}(3\mu-1)
+(4\gamma^{2}L^{2}-27\mu-12) \gamma L\coth\gamma
L \nonumber \\
&& -( 48+3\gamma^{2}L^{2}\left( \mu+4\right) -4\gamma^{4}L^{4})
\frac{\tanh\gamma L}{\gamma L}] \}. \label{S-zero}
\end{eqnarray}
The typical $\nu$-dependence of spectral density at zero frequency
is plotted in Fig.~6. We see a clear minimum at $\nu \simeq 3$. To
explain this minimum let us consider the resonant activation
phenomenon for this system.
\begin{figure}[htbp]
\begin{center}
\includegraphics[width=6cm]{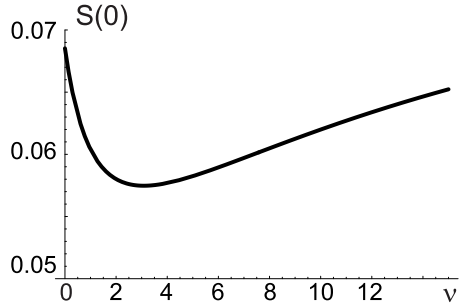}
\end{center}
\caption{\small \emph{Nonmonotonic behaviour of the spectral
density at zero frequency as a function of mean rate of flippings,
for the same parameters $L$, $a$, $D$ of Fig.~5.}} \label{beta_6}
\end{figure}
From the closed set of differential equations for MFPTs
$T_{+}\left( x\right)$ and $T_{-}\left(x\right)$ \cite{Bal}
\begin{eqnarray}
&& DT_{+}^{\prime\prime}-aT_{+}^{\prime}+\nu\left(
T_{-}-T_{+}\right)
=-1,\nonumber\\
&& DT_{-}^{\prime\prime}-aT_{-}^{\prime}+\nu\left(
T_{+}-T_{-}\right) =-1 \label{Han}
\end{eqnarray}
we calculate $T_{+}\left(  x\right)$ and $T_{-}\left( x\right)$,
\ie the MFPTs for positive $\eta\left( 0\right)=+1$ and negative
$\eta\left( 0\right)=-1$ initial value of the dichotomous noise,
with starting position of Brownian particles at the point $x$,
respectively. If we place the absorbing boundary at the point
$x=L$ we solve equations (\ref{Han}) with the following boundary
conditions: $T_{\pm}^{\prime}\left(  -L\right)  =0,\;
T_{\pm}\left( L\right) =0.$ The arithmetic average of MFPTs
$T\left(  x\right) =\left[ T_{+}\left(  x\right) +T_{-}\left(
x\right)  \right] \left/ 2\right.$ for initial position of
Brownian particles at the point $x=-L$ is
\begin{equation}
T\left( -L\right)  =\frac{4\nu L^{2}}{\gamma^{2}D^{2}}+\frac{a^{2}}%
{\gamma^{4}D^{3}}\left[  \cosh2\gamma L-1-\frac{\left(  \sinh2\gamma L-2\gamma
L\right)  ^{2}}{\cosh2\gamma L+\mu}\right]  .\label{Mean-Ar}%
\end{equation}

The behaviour of $T(-L)$ as a function of switchings mean rate has
a minimum, as shown in Fig.~7.
\begin{figure}[htbp]
\begin{center}
\includegraphics[width=6cm]{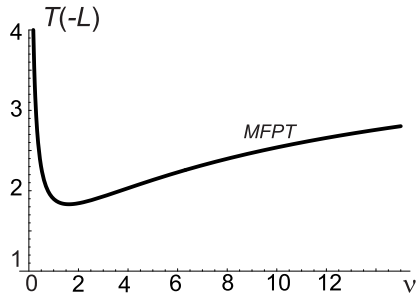}
\end{center}
\caption{\small \emph{Resonant activation phenomenon for MFPT
$T\left( -L\right) $. The parameters $L$, $a$ and $D$ are the same
as in Fig.~6.}} \label{beta_7}
\end{figure}
This effect was called in literature resonant activation: the
average residence time as a function of the barrier fluctuation
rate $\nu $ has a minimum at intermediate rates between very slow
and very fast fluctuations \cite{Doe}. In this range of rate $\nu
$, the crossing event is strongly correlated with the potential
fluctuations and Brownian particles overcome randomly switching
barrier in a minimal time. As a result, Brownian particle position
changes rapidly and very slow components of the random process
$x\left(t\right)$ are present in minor amounts: the spectral
density at zero frequency takes a minimum. Thus, the nonmonotonic
behaviour of the spectral density at zero frequency
$S\left(0\right)$ can be interpreted as a new characterization of
resonant activation phenomenon.

Finally we report in Fig.~8 the behaviour of spectral width $\Pi$
as a function of flippings mean rate $\nu$.
\begin{figure}[htbp]
\begin{center}
\includegraphics[width=5cm]{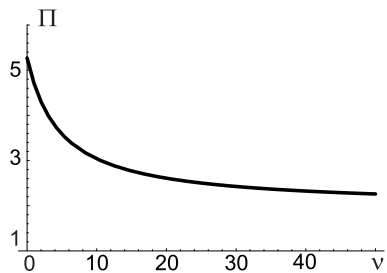}
\end{center}
\caption{\small \emph{Spectral width vs flippings mean rate for
the same parameters $L$, $a$ and $D$ of Fig.~5.}} \label{beta_8}
\end{figure}
We find a new nonlinear phenomenon: the spectral width decreases
with increasing mean rate of switchings, contrary to the linear
behaviour. As switchings mean rate increases, the slope of the
potential profile of Fig.~4 becomes less and less important. As a
result, the diffusion time of Brownian particle between the
reflecting boundaries $x=\pm L$ increases and is determined by a
free diffusion at very fast flippings. The random process $x\left(
t\right)$ therefore becomes more slow and the spectral width $\Pi$
decreases.

\section{Conclusions}

The exact formula for the spectral density of diffusion in
double-well potential for arbitrary noise intensity and arbitrary
parameters of potential profile was obtained. We found very rapid
narrowing of the spectrum with increasing height of a potential
barrier between steady states. We also derived the exact result
for spectral density of fluctuations in two-noise nonlinear
system, namely, for overdamped Brownian diffusion in randomly
flipping potential. We found a new characterization of resonant
activation phenomenon in the behaviour of spectral density at zero
frequency and new nonlinear effect associated with narrowing of
the spectrum of Brownian particle position with increasing mean
rate of switchings. Our analytical method enable us to investigate
more difficult problems as those with more complex potential
profiles.\ \\

This work has been supported by INTAS Grant 2001-0450, MIUR, INFM,
by Russian Foundation for Basic Research (project 02-02-17517), by
Federal Program "Scientific Schools of Russia" (project
1729.2003.2), and by Scientific Program "Universities of Russia"
(project 01.01.020).

\end{document}